\documentclass[journal-jpcc, manuscript=letter,layout=onecolumn]{achemso}

\setkeys{acs}{chaptertitle = true}

\usepackage{geometry}
\usepackage{enumerate}
\usepackage{graphicx,subfigure,mciteplus}
\usepackage{amssymb, amsmath}
\usepackage[flushleft]{threeparttable}
\usepackage{comment}
\usepackage{epstopdf, booktabs}
\usepackage{multirow,bigdelim}
\usepackage{ragged2e}
\makeatletter
\g@addto@macro\TPT@defaults{\footnotesize}
\makeatother
\usepackage{bm}
\usepackage{dcolumn}
\usepackage{bm}
\usepackage{longtable}
\usepackage{tabularx}
\usepackage{multirow}
\usepackage{ulem}
\usepackage[version=3]{mhchem}
\usepackage{color}
\usepackage[latin1]{inputenc}
\usepackage{array}
\usepackage{float}
\usepackage{leftidx}
\usepackage{placeins}
\usepackage[latin1]{inputenc}
\usepackage[T1]{fontenc}
\usepackage{multirow}
\usepackage{booktabs}
\usepackage{siunitx}
\usepackage{footnote}
\usepackage{dcolumn}
\usepackage{catchfilebetweentags}
\usepackage{xr-hyper}
\usepackage[nounderscore]{syntax}
\usepackage[version=3]{mhchem}
\usepackage{dcolumn}
\usepackage{siunitx}
\usepackage{amsmath}
\usepackage{amsthm}

\usepackage{hyperref}
\hypersetup{colorlinks,
	citecolor=blue
}
\usepackage{cleveref}

\mciteErrorOnUnknownfalse

\usepackage{geometry}
\usepackage{graphicx,subfigure,mciteplus}
\usepackage{amssymb, amsmath}
\usepackage[flushleft]{threeparttable}
\usepackage{comment}
\usepackage{epstopdf, booktabs}
\usepackage{mathtools}
\usepackage{multirow,bigdelim}
\usepackage{ragged2e}
\makeatletter
\g@addto@macro\TPT@defaults{\footnotesize}
\makeatother
\usepackage{bm}
\usepackage{dcolumn}
\usepackage{bm}
\usepackage{longtable}
\usepackage{tabularx}
\usepackage{multirow}
\usepackage{ulem}
\usepackage[version=3]{mhchem}
\usepackage{color}
\usepackage{amsthm}
\usepackage[latin1]{inputenc}
\usepackage{float}
\usepackage{leftidx}
\usepackage{placeins}
\usepackage[latin1]{inputenc}
\usepackage{catchfilebetweentags}
\usepackage{xr-hyper}
\usepackage[nounderscore]{syntax}
\usepackage[version=3]{mhchem}
\usepackage{dcolumn}
\usepackage{siunitx}
\usepackage{hyperref}
\usepackage{amsmath}
\usepackage{hyperref}
\hypersetup{colorlinks,
	citecolor=blue
}
\usepackage{cleveref}
\mciteErrorOnUnknownfalse


\newcolumntype{L}{D{.}{.}{2,5}}

\newcolumntype{L}{D{.}{.}{2,5}}

\author{Manas Sajjan}
\affiliation{Department of Chemistry,  Purdue University, West Lafayette, IN 47907, United States}
\altaffiliation{Purdue Quantum Science and Engineering Institute,Purdue University, West Lafayette, IN 47907, United States}
\author{Rishabh Gupta}
\affiliation{Department of Chemistry,  Purdue University, West Lafayette, IN 47907, United States}
\altaffiliation{Purdue Quantum Science and Engineering Institute,Purdue University, West Lafayette, IN 47907, United States}
\author{Sumit Suresh Kale}
\affiliation{Department of Chemistry,  Purdue University, West Lafayette, IN 47907, United States}
\altaffiliation{Purdue Quantum Science and Engineering Institute,Purdue University, West Lafayette, IN 47907, United States}
\author{Vinit Singh}
\affiliation{Department of Chemistry,  Purdue University, West Lafayette, IN 47907, United States}
\altaffiliation{Purdue Quantum Science and Engineering Institute,Purdue University, West Lafayette, IN 47907, United States}
\author{Keerthi Kumaran}
\affiliation{Department of Physics and Astronomy, Purdue University, West Lafayette, IN 47907, United States}
\altaffiliation{Purdue Quantum Science and Engineering Institute,Purdue University, West Lafayette, IN 47907, United States}
\author{Sabre Kais}
\affiliation{Department of Chemistry,  Purdue University, West Lafayette, IN 47907, United States}
\altaffiliation{Purdue Quantum Science and Engineering Institute,Purdue University, West Lafayette, IN 47907, United States}
\email{kais@purdue.edu}
\title[An \textsf{achemso} demo]
  {Physics inspired quantum simulation of resonating valence bond states- a prototypical template for a spin-liquid ground state}

\keywords{American Chemical Society, \LaTeX}

\begin{document}







\begin{abstract}
Spin-liquids- an emergent, exotic collective phase of matter have garnered enormous attention in recent years. While experimentally, many prospective candidates have been proposed and realized, theoretically modeling real materials that display such behavior may pose serious challenges due to the inherently high correlation content of such phases. Over the last few decades, the second-quantum revolution has been the harbinger of a novel computational paradigm capable of initiating a foundational evolution in computational physics. In this report, we strive to use the power of the latter to study a prototypical model - a spin-$\frac{1}{2}$-unit cell of a Kagome anti-ferromagnet. Extended lattices of such unit cells are known to possess a magnetically disordered spin-liquid ground state. We employ robust classical numerical techniques like Density-Matrix Renormalization Group (DMRG) to identify the nature of the ground state through a matrix-product state (MPS) formulation.  We subsequently use the gained insight to construct an auxillary hamiltonian with reduced measurables and also design an ansatz that is modular and gate efficient. With robust error-mitigation strategies, we are able to demonstrate that the said ansatz is capable of accurately representing the target ground state even on a real IBMQ backend within $1\%$ accuracy in energy. Since the protocol is linearly scaling $O(n)$ in the number of unit cells, gate requirements, and the number of measurements, it is straightforwardly extendable to larger Kagome lattices which can pave the way for efficient construction of spin-liquid ground states on a quantum device.
\end{abstract}

\section{Introduction}
Quantum spin liquids are exotic phases of matter that arise in systems of interacting quantum spins.\cite{broholm2020quantum} Unlike conventional magnets, where the spins align in a well-defined pattern at low temperatures, spin liquids do not exhibit long-range magnetic order even at absolute zero. Instead, they are characterized by a unique combination of long-range quantum entanglement, fractionalized excitations, \cite{han2012fractionalized} and emergent gauge fields \cite{knolle2019field, broholm2020quantum}. The strong entanglement between spins over long distances is responsible for the global topological order observed in spin liquids, which is distinct from the broken symmetries seen in conventional ordered states. 

Spin liquids can either have a gapless or a gapped excitation spectrum. Gapped spin liquids are well characterized by the global topological structure of their ground-state wave functions. They exhibit emergent quasi-particle excitations that have nontrivial statistical interactions for instance, anyonic statistics in two-dimension.\cite{wilczek1982quantum} Gapless spin liquids, on the other hand, are more challenging to characterize as their ground states can be highly degenerate, and even quasi-particle description completely breaks down in some cases. \cite{hermele2004stability} Gapless spin liquids exhibit power-law decay of correlation profiles of observable quantities.


Among various spin liquid systems, the antiferromagnetic Heisenberg model on the Kagome lattice holds particular significance. The two-dimensional lattice consisting of hexagons with corner-sharing triangles (See Fig. \ref{fig:kagome_cell}) introduces strong geometric frustration that leads to intriguing physics. Numerical simulations have suggested the existence of various spin liquid states on the Kagome lattice, including resonating valence bond (RVB) state and the Z2 spin liquid.\cite{liao2017gapless, han2012fractionalized, wang2018quantum}. {\color{black} Such RVB states were first investigated by Pauling \cite{malisoff1941nature} for describing $\pi$ bonded organic molecules and then subsequently used by Anderson \cite{anderson1973resonating} in connection to Mott insulators}. For extended lattices the presence of an RVB ground state usually implies that the valence bond pairs are itinerant and can delocalize throughout the lattice.

Experimental studies have identified several materials as potential hosts of spin liquid behavior on the Kagome lattice. For instance, Herbertsmithite $(ZnCu_3(OH)_6Cl_2)$, a mineral compound with copper ions arranged in a Kagome lattice, has shown signatures of spin-liquid behavior  \cite{norman2016colloquium, lee2007quantum, gregor2008nonmagnetic}. Another mineral is Barlowite $(Cu_4(OH)_6FBr)$, which exhibits a Kagome lattice formed by copper ions. \cite{han2014barlowite} 
Furthermore, $Na_4Ir_3O_8$ is a material where the Ir ions, having a spin-1/2, reside on a Hyperkagome 3D lattice. Studies suggest that the ground state of the Heisenberg model on this lattice may exhibit a quantum spin liquid phase with a spinon Fermi surface \cite{lawler2008gapless}. {\color{black} Such spinon excitations arise due to itinerant unpaired electrons not participating in the ground state bonding framework. The list extends to other polycrystallites like $(CsX)Cu_5O_2(PO_4)_2$ \cite{winiarski2019cs} and even in $Tb_3Sb_3Mg_2O_{14}$ where each of $Tb^{+3}$ and $Sb^{+5}$ have alternating Kagome sub-lattices\cite{https://doi.org/10.1002/pssb.201600256} or in other pyrochlore derivatives like in $La_3Sb_3Zn_2O_{14}$ \cite{sanders2016re3sb3zn2o14, chamorro2020chemistry}}. It's important to note that the formation of a perfect Kagome lattice in natural materials can be challenging due to factors like lattice distortions, impurities, and disorder. Despite the emergence of promising candidates, experimentally confirming the existence of a pristine spin liquid material remains a significant challenge. 

Despite the progress in experimental investigations, theoretical modeling of Kagome spin liquids remains challenging due to the strong correlations involved. In this paper, we focus on the theoretical investigation of a spin-1/2 anti-ferromagnetic Heisenberg model on the Kagome lattice. A potential initial step to understanding the Kagome lattice involves preparing its ground state and subsequently conducting investigations using the prepared ground state. This paper focuses on achieving the ground state of Kagome lattices using  Variational Quantum Eigensolver (VQE). Such quantum simulations, variational and otherwise, have become quite popular in recent years and been performed for a variety of other systems using diverse methodologies \cite{sajjan2022quantum, sajjan2021quantum, sureshbabu2021implementation, sajjan2023imaginary, sajjan2022magnetic, gupta2022variational, gupta2021maximal, gupta2021convergence, gupta2023hamiltonian, kale2021constructive, kale2020spin}. However, even constructing the ground state of a single plaquette within the Kagome lattice requires 12 qubits. Various studies \cite{mcclean2018barren, Wang_2021} have demonstrated that when attempting to attain the ground state of Hamiltonians with a significant number of qubits (>8), any Hamiltonian-agnostic ansatz, beyond a certain depth, inevitably leads to a barren plateau. This barren plateau phenomenon is attributed to the excessive expressibility of the ansatz \cite{PRXQuantum.3.010313, cerezo2021cost, larocca2022diagnosing}. To circumvent the issue, it is necessary to employ an ansatz that possesses just enough expressibility to capture the system's ground state, without being overly expressive for other arbitrary states.

In this study, we develop a physics-inspired ansatz that is capable of capturing the ground state of a single plaquette system. In order to achieve this, we generate 2-point correlation profiles between various pairs of lattice sites using Density-Matrix Renormalization Group (DMRG), which is computationally cheaper than exact diagonalization. We choose an ansatz that also conforms to the same correlation profile based on which pairs of lattice sites exhibit a substantial correlation. One can even go one step further to pick ansatz which has the same sign of the correlations as the ones obtained via DMRG. {\color{black} Furthermore, we also define two auxillary Hamiltonians, which can be treated as reduced version of the original Hamiltonian. Through appropriate classical pre-processing we show that the ground state of the auxillary Hamiltonians is identical to the ground state of the target Hamiltonian. By utilizing the reduced Hamiltonian, we can achieve certain benefits when simulating on the real device compared to using the original Hamiltonian. Specifically, the reduced Hamiltonian significantly diminishes the number of gates required for computing expectations of the Hamiltonian multiple times during the optimization process. This reduction in number of measurements enhances the performance of the VQE within the present era of Noisy Intermediate-Scale Quantum (NISQ) technologies. Naturally all our results will be validated by performing calculations resulting in high fidelity ground states on an actual quantum hardware}. 


{\color{black} Even though there exists another work which attempts to perform a digital quantum simulation of the Kagome unit cell using Hamiltonian variational ansatz\cite{PhysRevB.106.214429}, however unlike us, the said report do not perform computations on an actual hardware but merely provides results using noise models on a simulator. Also being inspired from a classical pre-processing based on a thorough DMRG analysis, the approach we follow and the ansatz we use is quite different from them. The modular two-qubit ansatz which Ref \cite{PhysRevB.106.214429} employs requires the use of two single-qubit gates and one $fSim$ gate which in turn requires a SWAP gate 
to be implemented (see Section IV in Ref \cite{PhysRevB.106.214429}). Each such SWAP gate would need on an average three CNOT gates for implementation (sequential arrangement of such CNOT gates will lead to some fortunate cancellation which motivates the discussion of an average case). As we shall see later, our ansatz is also modular with two-qubit connectivity requiring just a single CNOT gate. This ensures crisp convergence to the ground state even on the real hardware with a statistical error being $\le 1\%$. Besides, as we shall discuss, our approach would also linearly scale in the number of such two-qubit gates even for an extended Kagome lattice with multiple plaquettes making it efficient and less susceptible to noise injected from NISQ platforms. We shall also present a systematic analysis for the choice of optimizers, initial parameterization and error-mitigation strategies to use which is not explored elsewhere. Analog simulation of Kagome ground state has also been undertaken using the 219-atom programmable quantum simulator\cite{semeghini2021probing}. In this  approach, arrays of $^{87} \rm{Rb}$ atoms were placed on the edges of a Kagome lattice. Such atoms are excitable to a high-energy Rydberg state thereby turning on the dipolar interactions with neighboring vertices and effectively mimicking the dimer bonds in a RVB. Rydberg blockade naturally limits the density of such dimeric bonds on the atomic assembly at appropriate filling fraction. The assembly was adiabatically time evolved to create quantum states which inhabits the span of several paired valence bond configurations with no local order. The onset of a quantum spin liquid phase was detected by using both diagonal and off-diagonal string operators. Since the present work is a digital simulation, it is naturally apparent that our approach is entirely different from the aforesaid attempt but shall act as a complementary recipe.}


In the following section we investigate the Kagome unit cell classically using DMRG as a precursor for choosing a suitable ansatz. Thereafter we discuss the choice of ansatz and methods to define an auxillary hamiltonian to afford measurement reduction. We then present a discussion on the choice of optimizers, initial parameterization and error-mitigation strategies that shall be used and present the results on the actual hardware. We conclude thereafter discussing the extendability of our ansatz for larger systems and using recent experimental efforts to study myriad applications of RVB in physics and chemistry even beyond the precincts of spin-liquids. This highlights the importance of our investigation.

\section{The model}\label{sec:model}

\begin{figure}[htbp]
  \centering
  \includegraphics[width=0.6\textwidth]{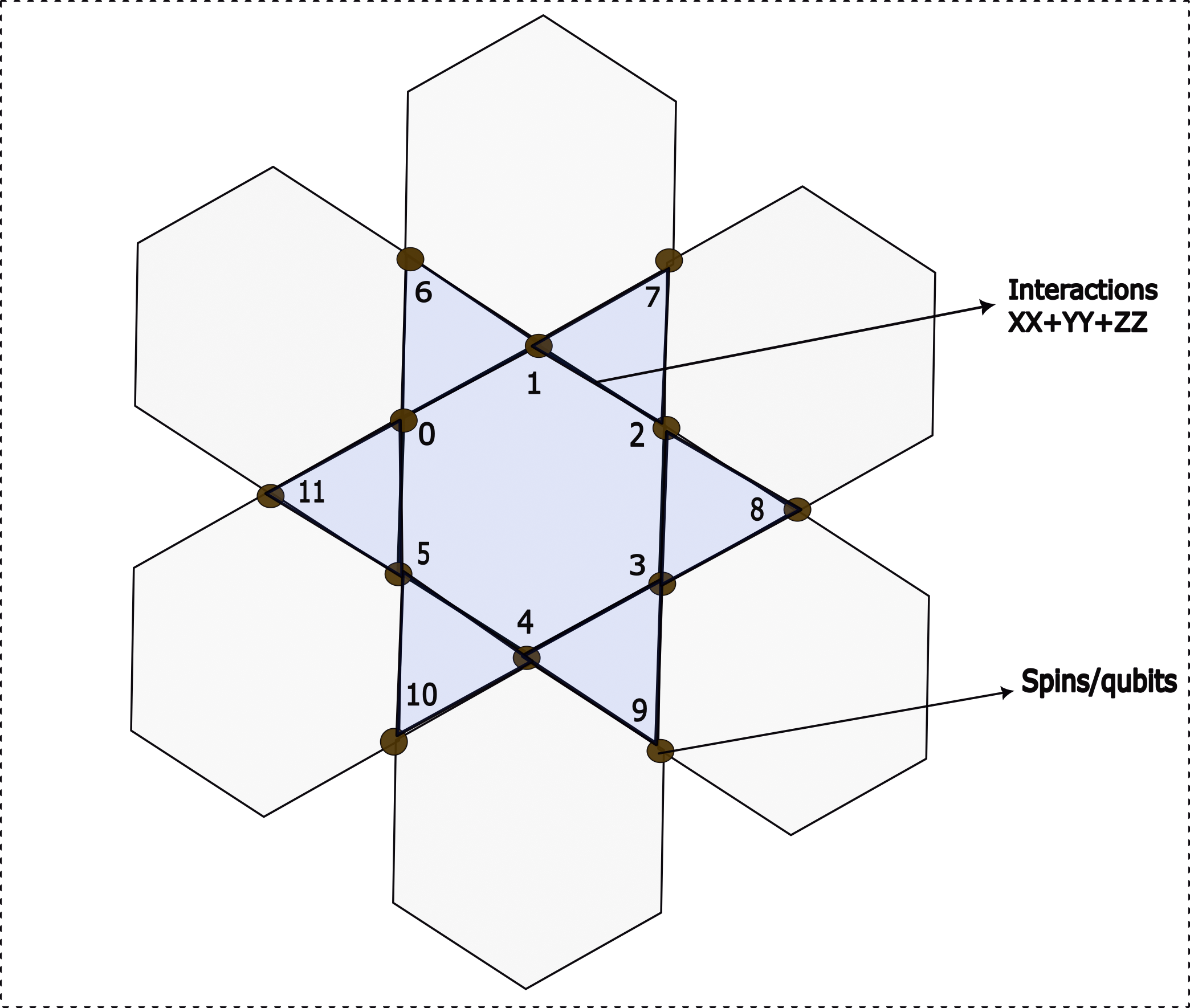}
  \caption{The unit cell (highlighted in blue) as used in this work within a larger Kagome lattice. Nodes of the unit cell (brown) are the location of spins/qubits. Displayed is the specific connectivity pattern as used in this work and also the contribution from each edge to the overall hamiltonian in Eq.\ref{full_H}.}
  \label{fig:kagome_cell}
\end{figure}

An anti-ferromagnetic Kagome lattice  consists of several tesselated unit cells, each of which can be described as being composed of qubits interacting according to the following Heisenberg Hamiltonian with homogeneous interactions (XXX)
\begin{eqnarray}
H(J)&= \sum_{(i, j) \in \eta} J(X_i X_j + Y_i Y_j + 
Z_i Z_j) \label{full_H}
\end{eqnarray}
where $(X,Y,Z)$ are usual Pauli operators and the indices $(i,j)$ refers to respective sites/qubits. The set $\eta$ is defined according to the adjacency matrix (connectivity pattern) of the interaction graph illustrated in Fig.\ref{fig:kagome_cell}. A pair of qubits (say $(k,m)$) sharing a specified edge in the unit cell shown in Fig.\ref{fig:kagome_cell} contributes a 
$J(X_k X_m + Y_k Y_m + Z_k Z_m)$
The connectivity graph we followed in this work, defines the set $\eta$ as 
\begin{eqnarray}
    \eta &= \{(0,1), (1,2), (2,3), (3,4), (4, 5), (5,0),(0,6), (1,6), (1,7), (2,7), (2,8), \\ \nonumber &(3,8), (3,9), (4,9), (4,10), (5,10), (5,11), (0,11)\} \label{eta_list}
\end{eqnarray}

The model can be interpreted as consisting of a flip-flop term 
$(X_i X_j + Y_i Y_j)$ which contributes to off-diagonal entries in the configuration basis (eigenbasis of $\sum_i Z_i$) by coupling configurations that has the same number of overall excitations but only differ in the of orientation of the spins at the $i$ and $j$th site only. The remaining $Z_i Z_j$ term contributes to diagonal entries and is responsible for primarily making the model anti-ferromagnetic given that the overall scale of the interaction strength $J \ge 0$. The model
possesses a set of discrete $\mathbf{Z}_2$ symmetries over reflection across the planes $xy, yz, xz$ which keeps the Hamiltonian defined in Eq.\ref{full_H} invariant under transformation of $X_i \mapsto -X_i$, $Y_i \mapsto -Y_i$ 
and $Z_i \mapsto -Z_i$ $\forall$ $i$. Besides the model also has a continuous $U(1)$ which leads to particle number conservation making $\sum_i Z_i$ commute with the hamiltonian in Eq.\ref{full_H}. In other words, the Hamiltonian matrix in the configuration basis would be blocked wherein within each block one shall only have configurations that belong to a particular eigenspace of $\sum_i Z_i$ tagged by a given eigenvalue. All such configurations have fixed number of excitations and the dimension of such a subspace having $k$ excitations would be ${N \choose k}$ where $N$ is the total number of spins. We shall see that the ground state which we are interested in would belong to $\langle \sum_i Z_i \rangle =0$ sector.

\section{Classical analysis: Motivation for choice of ansatz}\label{sec:ansatz_motivation} 

To motivate the choice of a reasonable ansatz that can reduce the computational time of actual quantum hardware without compromising accuracy we shall first study the unit cell classically. The technique of choice is DMRG using a Matrix-Product state ansatz. DMRG was first introduced by \cite{white1993density} as a method for studying one-dimensional quantum systems. The method has been routinely used ever since for analyzing strongly correlated electronic as well as spin systems with high accuracy\cite{}. The algorithm is based on the idea of truncating the Hilbert space of the system while preserving its essential properties. 
The specific version of the algorithm we use represents the target state of the system as a Matrix Product State (MPS).
In such an MPS ansatz, each component of the state of the system is written a tensor network comprising product of individual tensors defined at a local site in the system. The bond indices among the tensors is contracted while keeping the physical indices free. Mathematically an MPS can be described as a sum of the product of matrices, 

$$|\psi\rangle = \sum_{\{\vec{s}\}} \sum_{\{\alpha\}} \Lambda_{s_1}^{\alpha_1} \Lambda_{s_2}^{\alpha_1, \alpha_2} \cdots \Lambda_{s_{n-1}}^{\alpha_{n-1}, \alpha_{n}} \Lambda_{s_n}^{\alpha_{n-1}, \alpha_n} |s_1, s_2, s_3....s_n\rangle $$

where $\Lambda_{s_i}$ are complex square matrices of order $\chi$ known as the bond dimension. $\alpha_i$ are the bond indices and $s_i$ are the physical indices. The bond dimension of the local tensor quantifies the entanglement between the sites of the system. MPS with a fixed bond dimension $\chi$ can approximate a quantum state residing in a $N$-qubit Hilbert space using only $O(N\chi^2)$ parameters. One can then use this ansatz and minimize the energy of the system by iteratively optimizing the expectation value of the Hamiltonian until a desired level of accuracy is reached.

Unlike in 1D, even though DMRG in 2D has an unfavorable scaling over the width of the lattice, a number of techniques have been developed over the recent years to ameliorate the issues for practical usage as described in
\cite{stoudenmire2012studying, hyatt2019dmrg}. This has made the algorithm more favorable than many other popularly used methods in condensed matter physics like analytical treatments using Bethe ansatz\cite{}, Conformal field theory \cite{} (both of which are not extendable beyond 1D), Quantum-Monte Carlo-based methods which often suffer from non-negativity of numerically obtained probability values due to the infamous sign-problem \cite{}. Indeed some of the benchmark studies on potential suspects of 2D spin-liquids stabilized by anisotropy \cite{PhysRevB.74.012407}, multi-spin interactions \cite{PhysRevLett.106.157202} were initiated using DMRG. In fact, DMRG has been used to study anti-ferromagnet Kagome lattices (an extended version of the exact system we are interested in)\cite{yan2011spin},\cite{PhysRevLett.101.117203} quite fruitfully. 

\begin{figure}[htbp]
  \centering
  \includegraphics[width=0.99\textwidth]{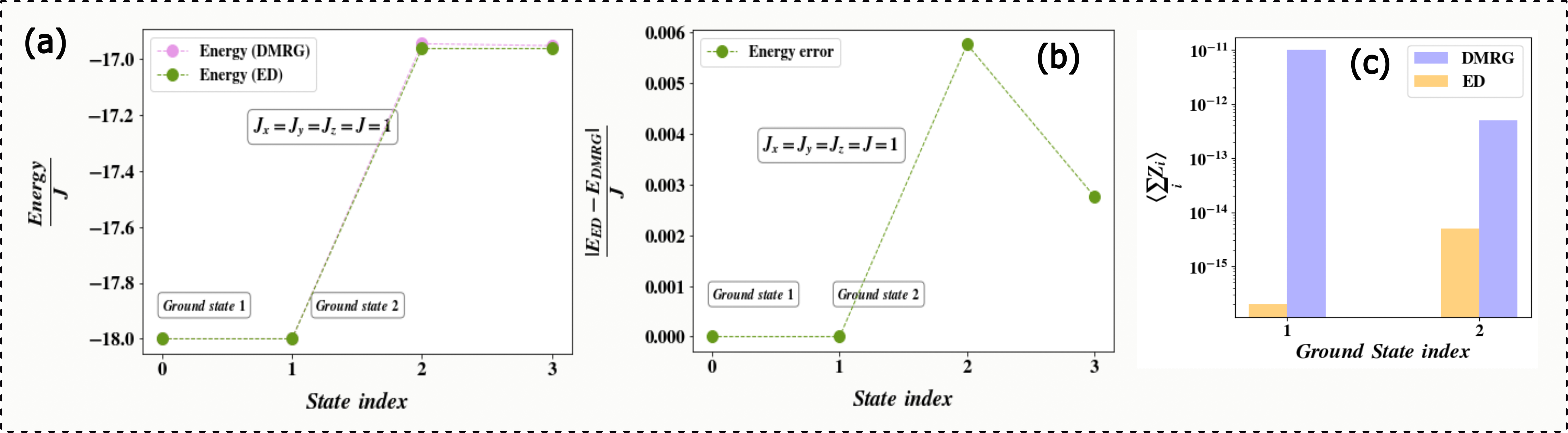}
  \caption{(a) The energy obtained from DMRG and exact diagonalization (ED) as described in the text for the 4 lowest energy eigenstates of Hamiltonian in Eq.\ref{full_H}. There are two degenerate ground states at -18.0J which are replicated by DMRG with near exact accuracy (states are marked as Ground state 1 and Ground state 2). (b)The energy error obtained from DMRG and exact diagonalization (ED) as described in the text for the 4 lowest energy eigenstates of Hamiltonian in Eq.\ref{full_H}. There are two degenerate ground states at -18.0J which are replicated by DMRG with near exact accuracy (states are marked as Ground state 1 and Ground state 2) (c) The $\langle \sum_i Z_i\rangle$ from the two ground state (See (a) or (b)) obtained from DMRG and ED. Both the states belong to $\langle \sum_i Z_i\rangle =0$ eigenvalue sector}
  \label{fig:en_DMRG_ED}
\end{figure}

Motivated by such developments we employ the algorithm to study classically the unit cell as described above. We use the Julia version of ITensor library \cite{} for all computations and employ an MPS with a maximum bond dimension of 200 and a maximum number of sweeps kept at 40 during the execution of the protocol. Arnoldi's method \cite{} is used for diagonalizing the reduced density matrices on the local Hilbert spaces. Spectrum of the latter was subsequently used to truncate the wavefunction MPS. To assess the quality of the numerical computation we compare all results with exact diagonalization (note that exact diagonalization is not required for our protocol, we just use it as a benchmark to show our DMRG results are in great agreement. For larger systems such a comparison need not be done). 

We compute the four lowest energy eigenstates of each from both methods and display the resulting energies in Fig.\ref{fig:en_DMRG_ED}(a-b). We see for all four states the energy error is less than 2\% of the exact values with the error being near zero for the two ground states we are interested in. Besides in Fig.\ref{fig:en_DMRG_ED}(c) we plot the $\langle \sum_i Z_i\rangle$ value for the two ground states as obtained from DMRG and ED and show that within numerical precision, both the ground states inhabit the $\langle \sum_i Z_i\rangle =0$ eigenvalue sector.

Next to assess the structural attributes of the MPS wavefunction and the exactly diagonalized one we compute the Neel correlation function defined as follows:
\begin{eqnarray}
\langle S_\alpha (k) S_\alpha (i) \rangle   \:\:\: \forall \alpha \in \{x,y,z\}, \:\:\:(k,i) \:\:\in \:\: \eta \label{eq:Neel_corr}
\end{eqnarray}

where $S_\alpha$ are the spin matrices defined in terms of Pauli operators $(X,Y,Z)$ as 
\begin{eqnarray}
S_{x} =\frac{X}{2}, & \nonumber
S_{y} =\frac{Y}{2}, & 
S_{z} =\frac{Z}{2} \nonumber
\end{eqnarray}

\begin{figure}[htbp]
  \centering
  \includegraphics[width=0.93\textwidth]{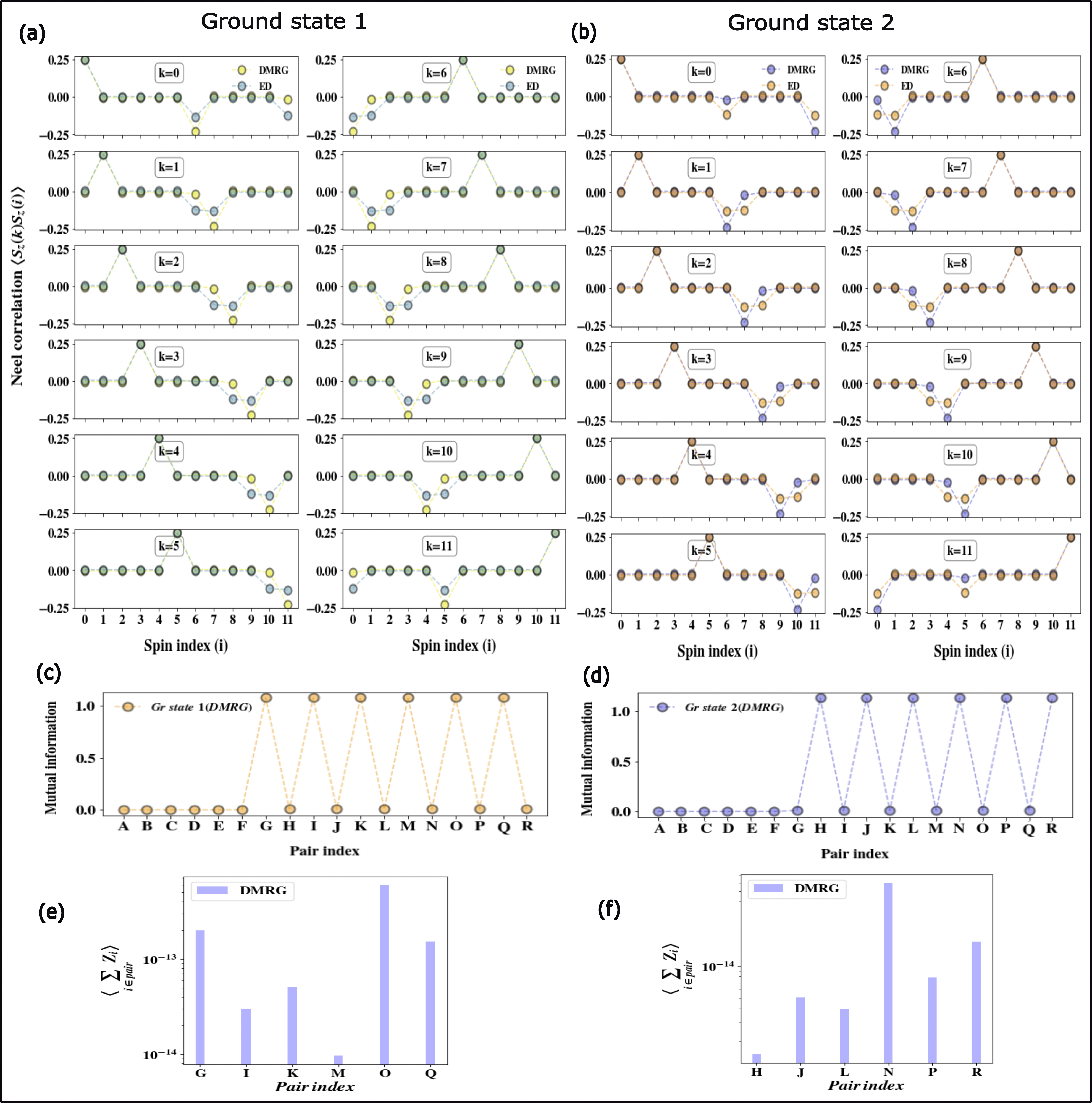}
  \caption{The Neel correlation function for all 12 spins in the unit cell defined in Fig.\ref{fig:kagome_cell} obtained from the \underline{first ground state} of both DMRG and ED as a function of spin indices along the unit cell. The reference spin is denoted as $k$ and all other target spins as $i$ consistent with the notation in (see Eq.\ref{eq:Neel_corr}). From the DMRG plots, we see prominent anti-correlation between certain pairs of spins only namely $(0,6)/(6,0), (1,7)/(7,1), (2,8)/(8,2), (3,9)/(9,3), (4,10)/(10,4), (5,11)/(11,5)$.The slight discrepancy between ED and DMRG stems from the fact that it is possible for numerical algorithms to return an arbitrary superposition of the two ground states seen in Fig.\ref{fig:en_DMRG_ED}. (b) Same as in (a) but for the \underline{second ground state} where we see prominent anti-correlation between certain pairs of spins only namely $(0,11)/(11,0), (1,6)/(6,1), (2,7)/(7,2), (3,8)/(8,3), (4,9)/(9,4), (5,10)/(10,5)$ (c)Plots for pair mutual information in the joint two-body state of all pairs (A-R) (see list Eq.\ref{eq:pair_list} for the definition of pairs)for both the ground states obtained from DMRG calculations. In ground state 1 (Gr state 1), the joint two-body state of the qubit pairs $G,I,K,M,O,Q$ have inherent quantum correlation (given the overall full MPS is a pure state) and hence possesses an entropy $S$ which cannot be reproduced by the sum of the entropies of the corresponding one-body reduced density matrices of the qubits. All other pairs seem to have a factorizable state. (d) Same as in (c) but for the second ground state where $H,J,L,N,P,R$ have inherent quantum correlation. (e) $\langle \sum_i Z_i\rangle$ where $i \in$ pairs $G,I,K,M,O,Q$ only indicating that such pairs shares a joint state with $\langle \sum_i Z_i\rangle =0$ in first ground state. (f) Same as in (e) but for pairs $H,J,L,N,P,R$ in second ground state indicating such each pair share a joint state with $\langle \sum_i Z_i\rangle =0$ }
  \label{fig:Neel_state_1}
\end{figure}

We display the results in Fig.\ref{fig:Neel_state_1}(a-b) for $\alpha=z$ i.e. with $S_z$ operators  
for each of the two ground states seen in Fig.\ref{fig:en_DMRG_ED} respectively. We see astonishing anti-correlation between specific pair of spins which are $(0,6)/(6,0), (1,7)/(7,1), (2,8)/(8,2)$, $(3,9)/(9,3), (4,10)/(10,4)$, $(5,11)/(11,5)$ in the first ground state from DMRG and between pairs $(0,11)/(11,0)$, 
$(1,6)/(6,1)$, $(2,7)/(7,2),(3,8)/(8,3),(4,9)/(9,4), (5,10)/(10,5)$ in the second ground state. It must be emphasized that for each reference spin $k$ we see a spike in the Neel correlation for $i=k$ too which is exactly 0.25. This is expected as $S_z (k)^2=\frac{1}{4}\mathcal{I}$. The orthogonality of the two obtained ground states dictates the correlation profiles of each peak on different sites leading to two sets of pairs as mentioned above. The states from ED for both the ground state seems to be such a superposition of the two degenerate ground states from the ones obtained from DMRG and hence shows anti-correlation for both sets of pairs above. The profiles for $\langle S_x (k) S_x (i) \rangle$ and $\langle S_y (k) S_y (i) \rangle$ are similar to the ones for $S_z$ due to the directional symmetry in the Hamiltonian in Eq.\ref{full_H}. Based on this correlation profile we are inclined to conclude that a valid representation of the ground state of the target Hamiltonian will be one wherein specific pair of spins have highly non-trivial interaction leading to anti-correlation. 

We then move on to analyze the mutual information $\mathcal{I}(i,j)$ which is defined as follows
\begin{eqnarray}
\mathcal{I}(i,j) = S(^1\rho(i)) + S(^1\rho(j))- S(^2\rho(i,j)) 
\end{eqnarray}
where $(i,j)$ denotes a pair in the set $\eta$ (in Eq.\ref{eta_list}) and the entropy is defined as $S(X)=-Tr(Xln(X))$ where $X$ is the reduced density matrix of the corresponding qubits. We designate the pairs in Eq.\ref{eta_list} using letters $A-R$ for convenience of representation. This encoding is defined as 
\begin{eqnarray}\label{eq:pair_list}
    & A=(0,1),
             B=(1,2),
             C=(2,3),
             D=(3,4), 
             E=(4, 5), \nonumber \\
            & F=(5,0), 
             G=(0,6), 
             H=(1,6), 
             I=(1,7), \nonumber \\
            & J=(2,7), 
             K=(2,8),  
             L=(3,8), 
             M=(3,9),  \nonumber \\
            & N=(4,9),  
             O=(4,10), 
             P=(5,10), 
             Q=(5,11), 
             R=(0,11) 
\end{eqnarray}

The results from the two ground-state MPS of DMRG calculations are displayed in Fig.\ref{fig:Neel_state_1}(c-d) as a function of a list of qubit-pairs (see Eq.\ref{eq:pair_list}). We see that for the first ground state, pairs like $G=(0,6), I=(1,7), K=(2,8), M=(3,9), O=(4,10), Q=(5,11)$ show unusually high mutual information but all other pairs have mutual information of zero. This is only possible if in this ground state (Gr-state-1) all other pairs except those above have factorizable reduced states but the two-qubits in each of the pairs $G,I,K,M,O,Q$ have inherently quantum correlated joint state $^2\rho(i,j)$ which is not factorizable (i.e. the correlation is such that the entropies of the one-particle reduced state of each qubit in the pair cannot reproduce the total joint entropy). Similar conclusion can be drawn for the other ground state (Gr-state-2) too but for pairs $H,J,L,N,P,R$ (see Eq.\ref{eq:pair_list}) for a definition of the pairs). Lastly in Fig.\ref{fig:Neel_state_1}(e-f) we plot the $\sum_{i \in D} \langle Z_i \rangle$ where the set $D=G,I,K,M,O,Q$ for the first ground state (see Fig.\ref{fig:Neel_state_1}(e)) and $D=H,J,L,N,P,R$ for the second ground state (see Fig.\ref{fig:Neel_state_1}(f)). This is done by computing the reduced density matrix $^2\rho(i,j)$ $\forall (i,j) \in D$ and then computing the said two body z-component of total spin average. The results indicate that the joint state of the spins in each such pair is such that not only it is factorizable from the remaining pairs (see Fig.\ref{fig:Neel_state_1}(e-f)) but also has a overall z-component of spin as zero. This is true for both the ground states even though the respective pairs where this happens are complementary.

The data from the DMRG simulations(see Fig.\ref{fig:Neel_state_1}(a-f), clearly states that the quantum states obtained from the MPS representation of DMRG calculations have special quantum correlation between specific pairs (either $G,I,K,M,O,Q$ or $H,J,L,N,P,R$ depending on which of the two ground state is investigated). All other pairs seems to have a factorizable state. Each of the correlated pairs shares a joint state which is made from exclusive amplitudes on the two-qubit basis configurations $|01\rangle$ and $|10\rangle$. This is 
bolstered by the fact that the Neel correlation in such pairs shows perfect anti-correlation (of magnitude $\frac{1}{4}$) and also the fact that the z-component of the total spin is zero. We will present a more direct evidence in the following section.

\subsection{Auxillary Hamiltonians with reduced measurements} \label{sec:Aux_H}

\begin{figure}[htbp]
  \centering
  \includegraphics[width=1.0\textwidth]{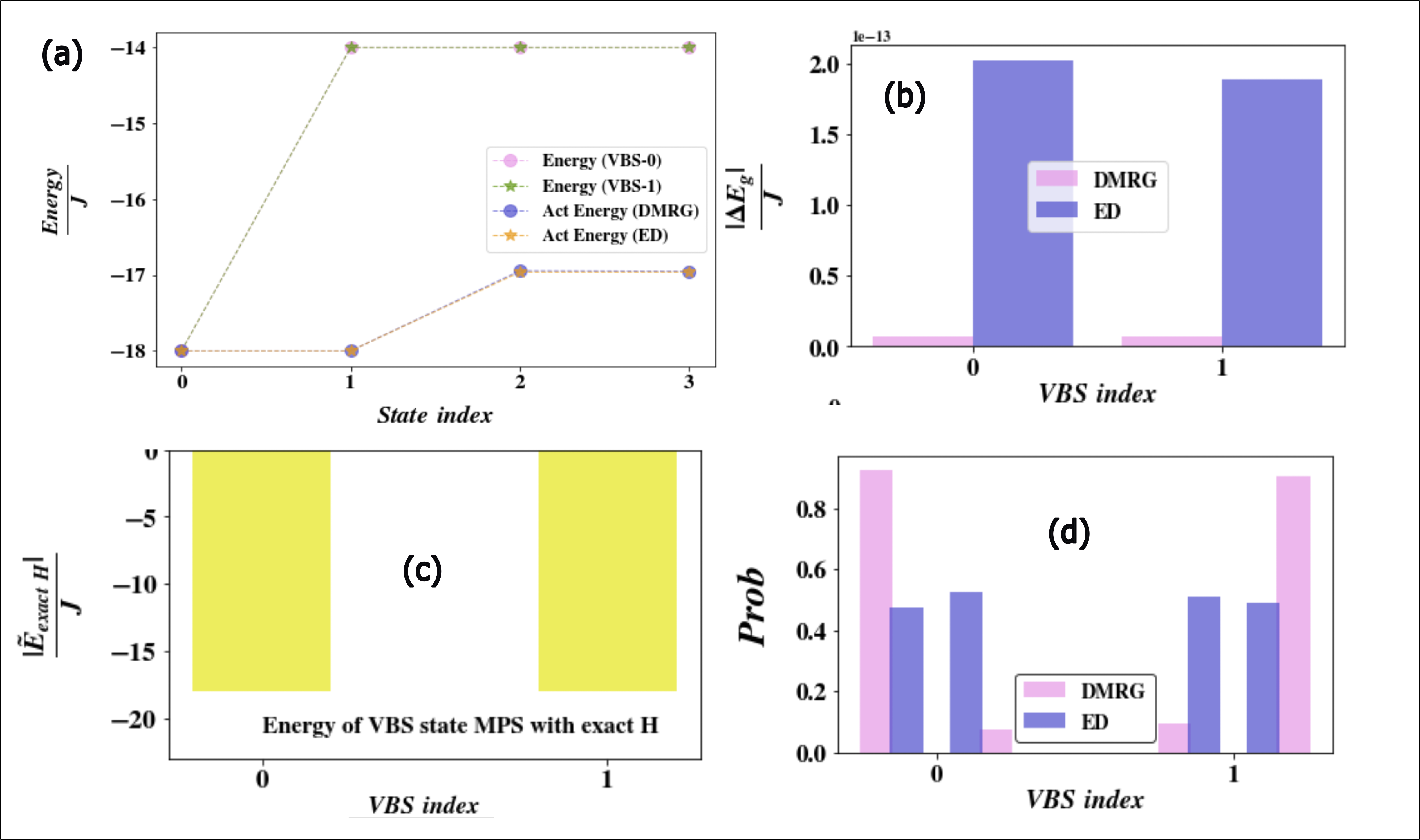}
  \caption{(a) Energy values of the four lowest energy eigenstates of $H_{VBS-0}(J)$, $H_{VBS-1}(J)$ and the full $H(J)$ in Eq.\ref{full_H}. All three Hamiltonians have iso-energetic ground states even though their excited states differ.(b) Energy error of the ground state energy values of $H_{VBS-0}(J)$, $H_{VBS-1}(J)$ from the ground state energy value obtained from DMRG/ED of $H(J)$ in Eq.\ref{full_H}. The ground state energy values exactly match. But this alone doesn't prove that they share the exact same ground state (it only establishes all three ground states are iso-energetic). (c) To prove that, we use the MPS of the ground state of $H_{VBS-1}(J)$ and $H_{VBS-0}(J)$ and compute its energy with respect to the full Hamiltonian in Eq.\ref{full_H} i.e compute $\langle \psi_{VBS-0}|H(J)|\psi_{VBS-0}\rangle$ and  $\langle \psi_{VBS-1}|H(J)|\psi_{VBS-1}\rangle$ where $H(J)$ is the full hamiltonian defined in Eq.\ref{full_H} and $\psi$ are the ground state of  $H_{VBS-1}(J)$ and $H_{VBS-0}(J)$. We obtain the same energy value as before. This establishes ground state of $H_{VBS-1}(J)$ and $H_{VBS-0}(J)$ act as a valid representation of the ground states of $H(J)$ in Eq.\ref{full_H} (d) We resolve each of the ground state of $H_{VBS-0}(J)$, $H_{VBS-1}(J)$ in the basis of the eigenstates (both DMRG/ED) of the hamiltonian in Eq.\ref{full_H}. We see for each the probability adds up to 1 indicating that $\psi_{VBS-0}$ and $\psi_{VBS-1}$ lives in the span of the ground state of $H(J)$ in Eq.\ref{full_H}.  }
  \label{fig:Energy_RVB_states}
\end{figure}

In this section, we show that the problem of finding the ground state of the target Hamiltonian in Eq.\ref{full_H} is equivalent to finding the ground state of two auxiliary Hamiltonians ($H_1, H_2$). The anti-correlation profile in the previous section coupled with the mutual information data indicates that special qubit pairs may share a non-factorizable joint state in the span of configurations $|10\rangle, |01\rangle$. In this section we engineer two simple hamiltonians whose ground state happens to be a perfect match with the ground state of the hamiltonian in Eq.\ref{full_H}. We call such hamiltonians as 
$H_{VBS-0}(J)$ and $H_{VBS-1}(J)$ anticipating the valence bond nature of its respective ground states\cite{PhysRevB.66.132408}. The interaction terms within the two hamiltonians are defined as  

\begin{eqnarray}
H_{VBS-0}(J)&= \sum_{(i, j) \in \kappa_1} J(X_i X_j + Y_i Y_j + 
Z_i Z_j) \label{H_VBS_1}  \\
H_{VBS-1}(J)&= \sum_{(i, j) \in \kappa_2} J(X_i X_j + Y_i Y_j + 
Z_i Z_j) \label{H_VBS_2} 
\end{eqnarray}

where the two sets $\kappa_1$ and $\kappa_2$ consist of pair of spins defined as follows: 
\begin{eqnarray}
    \kappa_1 = \{G:(0,6), I:(1,7), K:(2,8), M:(3,9), O:(4,10), Q:(5,11)\}  \nonumber\\
    \kappa_2 = \{R:(0,11), H:(1,6), J:(2,7), L:(3,8), N:(4,9), P:(5,10)\} \label{kappa_def}
\end{eqnarray}
We compute the four lowest energy eigenstates of both $H_{VBS-0}(J)$ and $H_{VBS-1}(J)$ and compare the results with ED and DMRG calculations of the full system in Eq.\ref{full_H} in Fig\ref{fig:Energy_RVB_states}(a). We see even though the two auxiliary Hamiltonians $H_{VBS-0}(J)$ and $H_{VBS-1}(J)$ differ from the full system in terms of excited state energy values, the ground state energy values of the $H_{VBS-0}(J)$ and $H_{VBS-1}(J)$ surprisingly match with the ground state energy value of hamiltonian in Eq.\ref{full_H}. This doesn't yet prove that all of these Hamiltonians have the same ground state, it only shows that $H_{VBS-0}(J)$ and $H_{VBS-1}(J)$ has an iso-energetic ground state with the Hamiltonian in Eq.\ref{full_H}. We verify this again in Fig\ref{fig:Energy_RVB_states}(b) we display the energy error of the $H_{VBS-1}(J)$ and $H_{VBS-0}(J)$ from the energy of Hamiltonian in Eq.\ref{full_H} indicating that the errors are actually quite low. 
In Fig\ref{fig:Energy_RVB_states}(c), we use the MPS of the ground state of $H_{VBS-1}(J)$ and $H_{VBS-0}(J)$ and compute its energy with respect to the full Hamiltonian in Eq.\ref{full_H} i.e compute 
$\langle \psi_{VBS-0}|H(J)|\psi_{VBS-0}\rangle$ and  $\langle \psi_{VBS-1}|H(J)|\psi_{VBS-1}\rangle$ where $H(J)$ is the full hamiltonian defined in Eq.\ref{full_H} and $\psi$ are the ground state of  $H_{VBS-1}(J)$ and $H_{VBS-0}(J)$. We see that the $\psi_{VBS-0}$ and $\psi_{VBS-1}$ also act as a valid choice of the ground state of hamiltonian in Eq.\ref{full_H} and return the exact same energy as before. By variational theorem, it is possible to then construct any valid ground states of Eq.\ref{full_H} from these two states and vice versa. To prove this assertion, we use as basis the two ground states obtained from DMRG and ED discussed before and resolve each of $\psi_{VBS-0}$ and $\psi_{VBS-1}$. We find that the probabilities of projecting any of the VBS state onto the two degenerate ground state obtained from DMRG/ED add upto 1. This is displayed in Fig\ref{fig:Energy_RVB_states}(d). This indicates that $\psi_{VBS-0}$ and $\psi_{VBS-1}$ lives in the span of the two eigenstates from DMRG and ED of the hamiltonian in Eq.\ref{full_H}. We thus have conclusively proved that the three hamiltonians- $H_{VBS-0}(J)$, $H_{VBS-1}(J)$ and the one defined as $H(J)$ in Eq.\ref{full_H} not only has an iso-energetic ground state  but actually shares the same ground state in the sense that ground state of one can be used as a valid representation for the ground state of the other two and lives in the span of the ground state of other two. 

We shall choose either Hamiltonian $H_{VBS-0}(J)$, $H_{VBS-1}(J)$ for all computations henceforth. The advantages of choosing either of the two are in the fact that both $H_{VBS-0}(J)$, $H_{VBS-1}(J)$ have reduced connectivity compared to hamiltonian in Eq.\ref{full_H} and for VQE computations, one needs reduced number of Pauli measurements for each (see list in Eq.\ref{kappa_def}). This saves computational load and mitigates error drastically as there are less number of expectation values for the noise to corrupt. Also, both hamiltonians have reduced density of states near the ground energy level (see Fig.\ref{fig:Energy_RVB_states}(a) which shows that the next excitation for either is at $-14J$) compared to the hamiltonian in Eq.\ref{full_H} so it is less likely for the VQE algorithm to be trapped locally and return a superposition of ground and higher energy states.

\subsection{Choice of ansatz}\label{sec:ansatz_choice}
\begin{figure}[h!]
    \centering
    \includegraphics[width=0.99\textwidth]{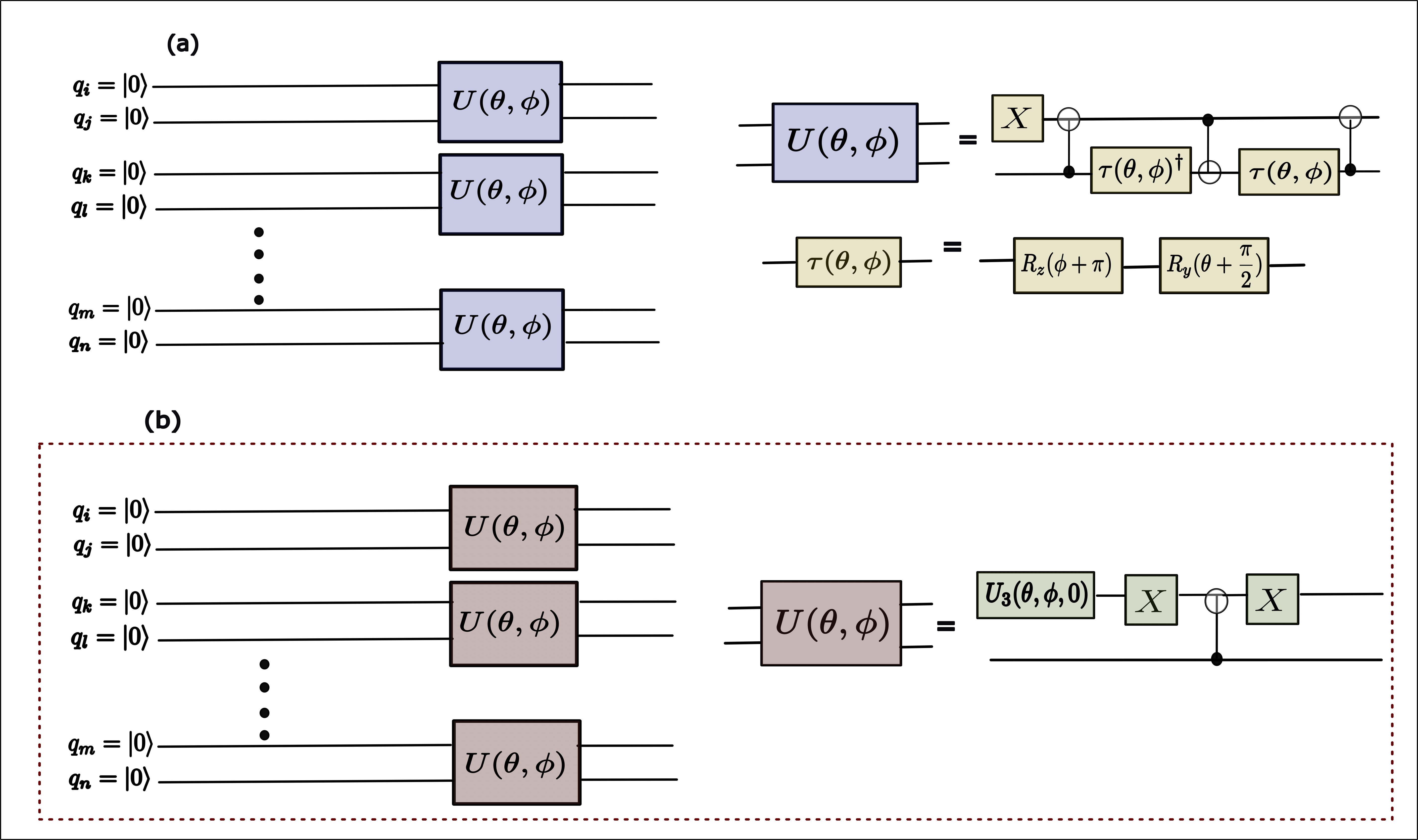}
    \caption{(a) A prospective choice of an ansatz which consist of a two-qubit  gadget $U(\theta, \phi)$ between pairs like $(i,j), (k,l), (m,n) \in \kappa_1$. The gadget implements a Gievens unitary rotation capable of generating a state in the single-excitation subspace i.e. in the span of $|10\rangle$ and $|01\rangle$ .The elemental gate decomposition of the gadget is given alongside. The number of unique parameters $(\theta, \phi)$ is 2 due to symmetry in the desired state as discussed before
    (b) The complete 12-qubit ansatz as used for the VQE calculations in this work. The two-qubit gadget $U(\theta, \phi)$ in this case again acts only between interacting pairs like $(i,j), (k,l), (m,n) \in \kappa_1$. The gadget comprises a single-qubit $U_3$ gate and a $CNOT$ gate controlled on state $|0\rangle$ of the control qubit. The unique parameter count is 2 and these parameters are denoted as $(\theta, \phi)$. They collectively define the operations of the $U_3$ gate in each couplet. The $CNOT$ in each pair creates a correlated two-qubit state in the span of $|10\rangle$ and $|01\rangle$. We discuss about such interacting couplets in next section specifically. Note that this ansatz consists of 1 CNOT gate per interacting pair which makes it more economical and less susceptible to gate infidelities compared to (a) thereby justifying its choice}
    \label{fig:ansatz_full}
\end{figure}

Based on the DMRG computations in the above section and the discussion in Section. \ref{sec:Aux_H} we can safely prepare the ground state of $H_{VBS-1}(J)$ and $H_{VBS-0}(J)$ and claim that those are ground states of the target hamiltonian defined in Eq.\ref{full_H}. As established from the DMRG calculations, the ground states of $H_{VBS-1}(J)$ and $H_{VBS-0}(J)$ are known to be singlet pairs (also the ground state of Eq.\ref{full_H} possesses properties satisfying this as seen in correlation profile before) with even a subspace $U(1)$ symmetry of a single excitation within each interacting pair, we propose to use either of two prospective ansatze as described in Fig.\ref{fig:ansatz_full} (a-b). Each of the two ansatz has a modular structure with exclusive connectivity between the following interacting pairs of spins. 
\begin{eqnarray}
    \kappa_1 = \{G:(0,6), I:(1,7), K:(2,8), M:(3,9), O:(4,10), Q:(5,11)\}
\end{eqnarray} \label{kappa_def}
and each of which can prepare ground state of $H_{VBS-0}(J)$.

Both the prospective candidates prepares a 
state $\psi(\theta,\phi)$ between each such interacting pair described as 
$\psi(\theta,\phi) = \cos(\theta/2)|01\rangle +$ $\sin(\theta/2)\exp(i\phi)|10\rangle$ in the above list. The joint many-body state of all the pairs will a kronecker product of such two-qubit states $\psi(\theta,\phi)$. The ansatz in Fig.\ref{fig:ansatz_full}(a) does this by starting from the initial configuration of $|01\rangle$ which is generated by the use of the first $X$ gate (little endian ordering is followed) in the inset of Fig.\ref{fig:ansatz_full}(a). Once such a configuration is created in the single excitation subspace, the two-qubit gadget $U(\theta, \phi)$ for this ansatz thereafter implements a Givens unitary ($G(\theta, \phi)$) \cite{PhysRevA.98.022322, arrazola2022universal, gard2020efficient, huggins2021efficient} defined as follows:
\begin{eqnarray}
G(\theta, \phi) &= 
\begin{pmatrix}
1 & 0 & 0 &0 \\
0 & \rm{cos}(\frac{\theta}{2}) & e^{i\phi} \rm{sin}(\frac{\theta}{2}) & 0\\
0 & e^{-i\phi}\rm{sin}(\frac{\theta}{2}) & \rm{cos}(\frac{\theta}{2}) & 0 \\
0 & 0 & 0 & 1
\end{pmatrix}
\end{eqnarray}
Thus the overall unitary is $U(\theta, \phi) = (X\otimes I) G(\theta, \phi)$ which creates the desired superposition. However as is seen in the inset of Fig.\ref{fig:ansatz_full}(a) wherein the full elemental decomposition of $G(\theta, \phi)$ is explicated, the ansatz requires 3 CNOT gates per interacting pair which enhances the likelihood of the implementation to be error prone on a near-term device. Contrary to this popular choice, we present in this manuscript another ansatz which is far more economical.
Like the one above, this ansatz also comprises of a pairwise interacting gadget $U(\theta, \phi)$ which only couples pair of spins $(p,q) \in \kappa_1$. The unique parameter count as before will be 2 and be denoted $(\theta, \phi)$. It is kept the same for all pairs due to similar correlation profiles, mutual information values and symmetry of each pair (see Fig.\ref{fig:Neel_state_1}). However unlike before, the ansatz in Fig.\ref{fig:ansatz_full}(b) (see inset) implements the target unitary as $U(\theta, \phi) = (U_3(\theta, \phi, 0) \otimes I)(X\otimes I) CNOT (X\otimes I)$ thereby requiring just 1 two-qubit gate ($CNOT$) for each pair. The single-qubit
$U_3(\theta, \phi, 0)$ gate creates the superposition of the control qubit thereby leading to the formation of a state in the span of $|00\rangle$ and $|10\rangle$. The $CNOT$ gate being controlled on the state $|0\rangle$ thereafter exclusively transforms  $|00\rangle \mapsto |01\rangle$. To use this scheme, one must emphasize that is mandatory to initialize the qubits in $|0\rangle^{\otimes n}$.
Henceforth, we shall use this ansatz for all computations in the manuscript.

\section{Results and Discussion}\label{sec:results_discussions}

\subsection{Choice of optimizer}

\begin{figure}[ht!]
  \centering
  \includegraphics[width=0.8\textwidth]{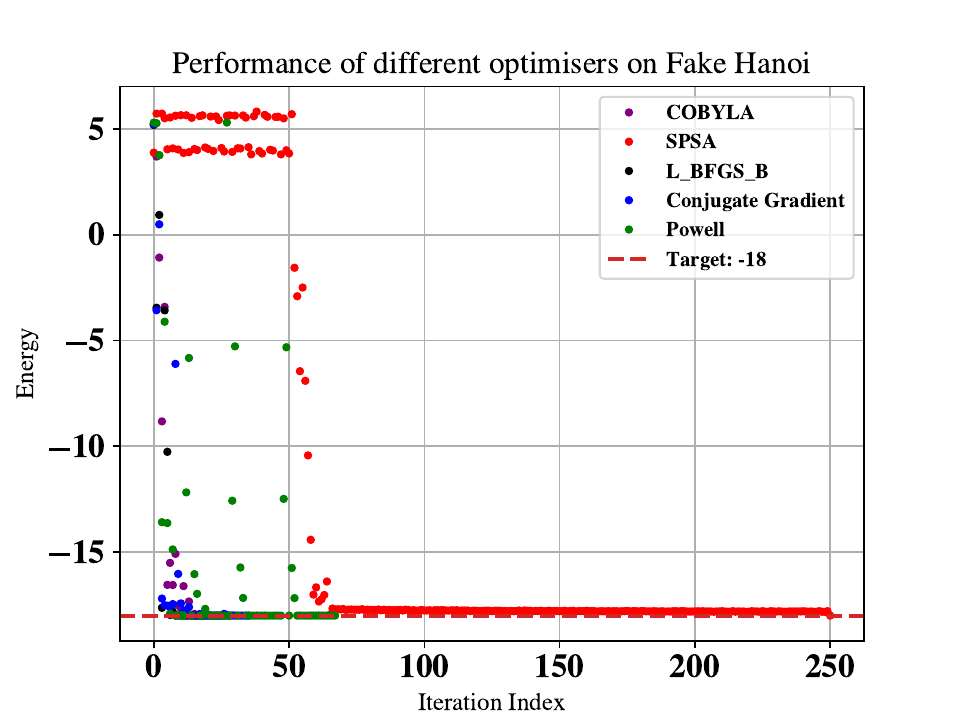}
  \caption{Performance of different optimizers when VQE computations are performed for the full 12-qubit ansatz by importing the noise model (\textit{FakeHanoi} backend) on \textit{ibmq-qasm} simulator.}
  \label{fig:optimzers}
\end{figure}

To analyze the performance of various optimizer we performed numerous calculations on \textit{ibmq-qasm} simulator by importing noise models (choice of the noise model is due to 
\textit{FakeHanoi} backend) using several available optimizer. The code base to perform all quantum simulations are implemented using Qiskit \cite{Qiskit}. Fig.\ref{fig:optimzers} shows the convergence of the  energy as a function of the epochs (iteration index) and it can be seen that most optimizer perform well on this ansatz. We see that while most optimizers perform reasonably well and eventually attains the desired accuracy, COBYLA provides the fastest convergence within 10-20 iterations. This has been corroborated multiple times for this specific case with different initial parameters. Since enhanced number of iterations on a quantum device would incur more error due to increased number of operations and decoherence of the qubit register, we stick to using COBYLA as the preferred optimizer for all cases henceforth.

\subsection{Choice of initial parameters and error-mitigation schemes using a single interacting component of the ansatz: a 2-qubit system} \label{intuition}

\begin{figure}[ht!]
  \centering
  \includegraphics[width=0.5\textwidth]{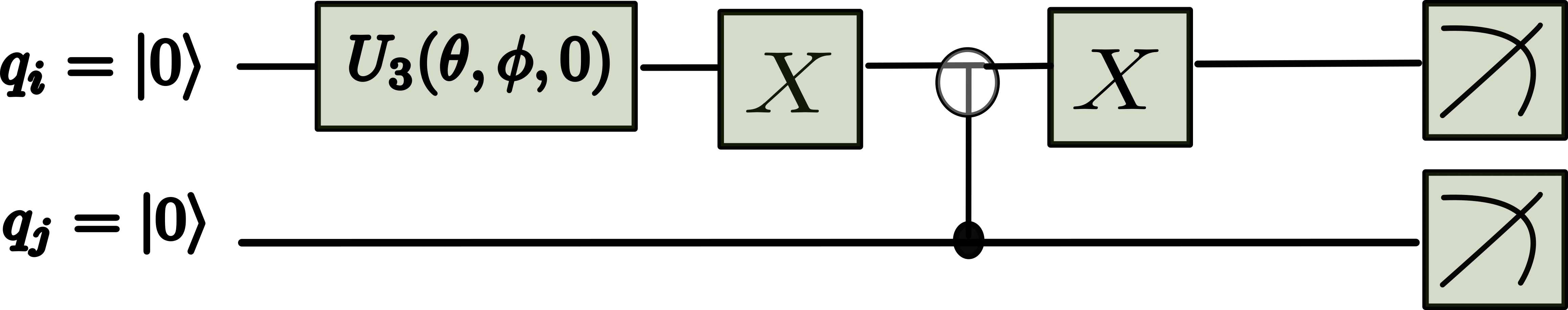}
  \caption{The 2-qubit ansatz used within the VQE calculations.}
  \label{fig:ansatz}
\end{figure}

Our analysis in Section \ref{sec:ansatz_motivation} revealed anti-correlation between 2 adjacent lattice sites inhabiting the triangular facets of the Kagome unit cell (See Fig.\ref{fig:kagome_cell}). Also the pinwheel structure of the Kagome lattice unit cell has $C_6$ symmetry. Therefore, starting with the anti-correlated 2-qubit system can provide a good foundation for comprehending the ground state of 12-qubit Kagome lattice. Based on this finding for a 2-qubit system, we propose an ansatz given by: $\psi(\theta,\phi) = \cos(\theta/2)|01\rangle + \sin(\theta/2)\exp(i\phi)|10\rangle$. The corresponding ansatz can be implemented onto a quantum circuit just by using a U3($\theta, \phi,0$) gate, and a CNOT gate as shown in Fig. \ref{fig:ansatz}.

\begin{figure}[ht!]
  \centering
  \includegraphics[width=0.99\textwidth]{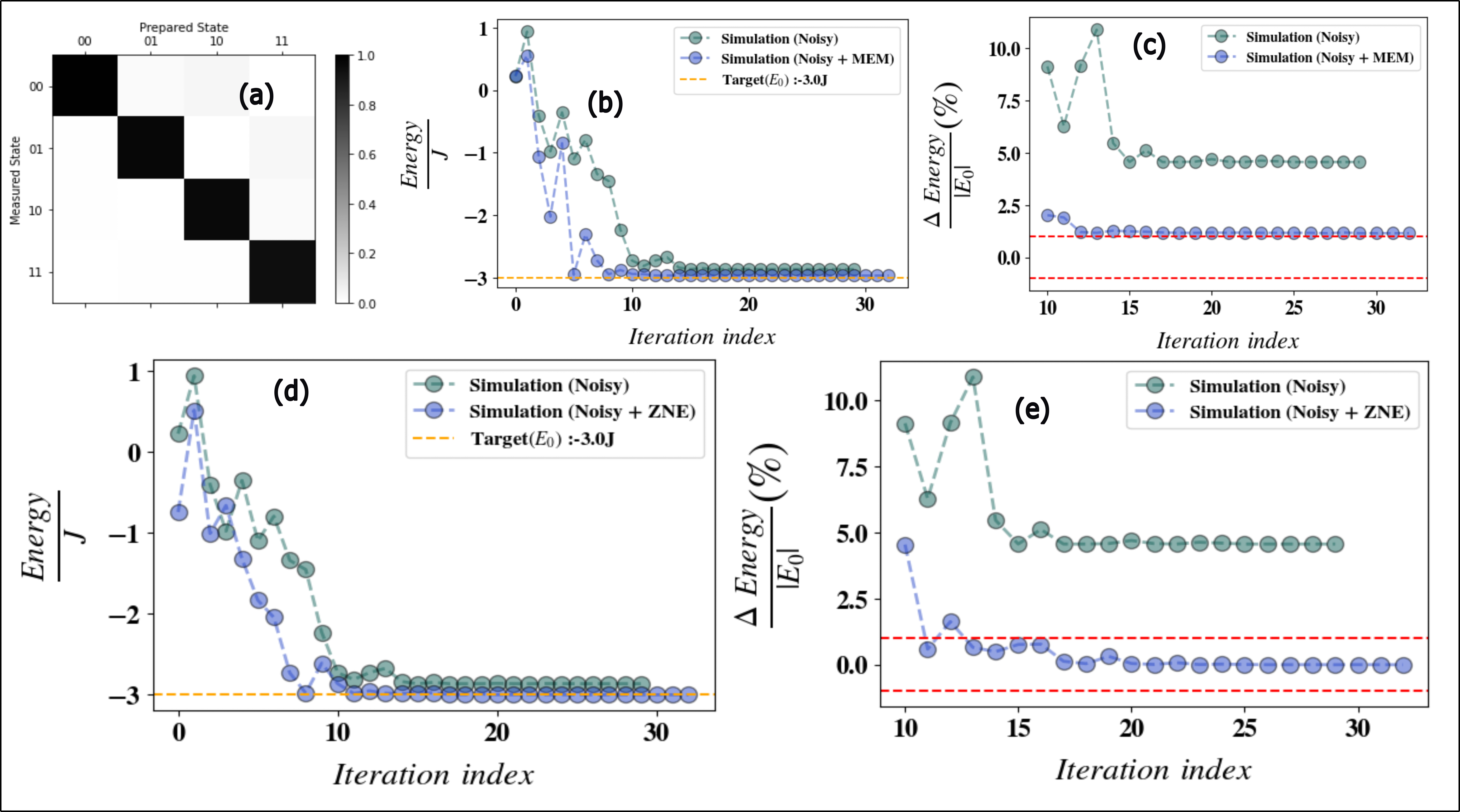}
  \caption{(a) The calibration matrix for all the 4 configurations of 2-qubit example (See Fig.\ref{fig:ansatz}) (b) Plots of energy, as a function of the number of epochs using a noise model coming from \textit{FakeHanoi} backend, with and without measurement error mitigation (MEM). (c) Same as in (b) except the y-axis is relative $\%$ energy error with respect to the true energy of $E_0=-3 J$, especially focusing on the last few iterations. The red dashed lines indicate $\%$energy error = $\pm 1 \%$. (c) Plots of energy, as a function of the number of epochs using a noise model coming from \textit{FakeHanoi} backend, with and without zero-noise extrapolation (ZNE). (e) Same as in (d) except the y-axis is relative $\%$ energy error in energy with respect to the true energy of $E_0=-3 J$ especially focusing on the last few iterations. The red dashed lines indicate $\%$energy error = $\pm 1 \%$. Unlike in (c) where MEM could barely attain an accuracy within $\pm 1\%$ error range, we see ZNE results are below the same quite effortlessly. All results are obtained by importing the said noise model on \textit{ibmq-qasm} simulator with COBYLA as the optimizer.}
  \label{fig:2_qubit_all_pics}
\end{figure}
Intuition about the performance of this 2-qubit ansatz can be beneficial for achieving convergence in the overall optimization algorithm for all 12 qubits, particularly when operating on real-world quantum devices, which are often noisy. We shall therefore use this 2-qubit ansatz to study what kind of initial parameters are to be used for warm-starting on a real quantum device for the 12 qubit case. We shall also study what should be the optimal choice of the error mitigation schemes. Incorporating such effective error mitigation techniques can significantly improve the accuracy of the final results. 
The 2-qubit hamiltonian for any one interacting pair (say (0,6)th qubit pair) is given by 
is 
\begin{eqnarray} \label{eq:2_qub_ham}
^2H = X_0 X_1 + Y_0 Y_1 + Z_0 Z_1
\end{eqnarray}
where we have re-labelled the dummy indices. The ground state of this hamiltonian is variationally obtained by minimizing the parameters ($\theta, \phi$) within the ansatz described in Fig.\ref{fig:ansatz}. To do this we also use several different choices of error-mitigation techniques by evaluating their effectiveness in our specific case against imported noise model (choice of the noise model is due to \textit{FakeHanoi} backend) in the \textit{ibmq-qasm} simulator. These schemes includes Zero-Noise Extrapolation (ZNE)\cite{temme2017error}, Pauli Twirling (popularly called T-Rex)\cite{van2022model}, and Measurement Error Mitigation (MEM) \cite{barron2020measurement}, which are the commonly used methods to mitigate errors in quantum computing. The code base to perform all such simulations with and without error-mitigation schemes are implemented using Qiskit \cite{Qiskit}. Various factors influence the performance of each of these techniques,
when it is employed to attenuate the effect of noise, including but not limited to the following. For instance,
ZNE can become increasingly complex as the number of qubits grows as well as the fidelity of quantum gates is high, as it requires running the same circuit multiple times with different levels of noise. In contrast, Pauli twirling can be applied more efficiently, as it involves preparing a more robust single noisy state that can be used for multiple operations. Furthermore, the MEM method builds a calibration matrix which is shown in Fig.\ref{fig:2_qubit_all_pics}(a). The size of this calibration matrix grows exponentially which therefore can be quite a costly exercise to evaluate exactly when the number of qubits becomes high. Furthermore, in the context of Variational Quantum Eigensolver (VQE), it is worth noting that the performance of different error mitigation schemes varies depending on the choice of optimizer used in the optimization process which in turn drastically affects the quality of convergence \cite{sung2020using}.

We use the 2-qubit hamiltonian defined in Eq.\ref{eq:2_qub_ham} and optimize the ansatz in Fig.\ref{fig:ansatz} with respect to the variational parameters using noise model imported from \textit{FakeHanoi} backend. The corresponding results for the energy function and the relative $\%$ energy error vs the epochs when using MEM are shown in Fig.\ref{fig:2_qubit_all_pics}(b-c). As can be seen in the Fig.\ref{fig:2_qubit_all_pics}(c), the $\%$ error for the case of \textit{FakeHanoi} backend without considering any error mitigation scheme is around 5-10$\%$. When we use the calibration matrix from MEM, the error is mitigated and comes down to around 1-2$\%$ (see Fig.\ref{fig:2_qubit_all_pics}(c)) but not below that. On the other hand if ZNE is used, the results as illustrated in Fig.\ref{fig:2_qubit_all_pics}(d-e) we see that it is possible to reach even below 1$\%$ error in the obtained final energy. Thus MEM was unable to reduce the error percentage as much as ZNE could do on the noisy model, and thus it will not be selected for executing VQE for the full 12-qubit system on an actual quantum device as we shall explore next. From the results of the computation with ZNE, the final converged parameter which minimized the energy for the 2-qubit case was found to be $\theta \sim \pi/2$ and $\phi \sim \pi$, which is essentially a Bell state of the kind $\frac{|10\rangle - |01\rangle}{2}$ thereby further corroborating our previous findings of having anti-correlated pairs. We use this set of parameters as an initial warm-start for the full 12-qubit case which we investigate next.

\subsection{Simulations on a real quantum hardware}

In this section, we show the results corresponding to the VQE calculation for finding the ground state energy of the 12-qubit Kagome lattice on the \textit{ibm$\_$hanoi} which is one of the IBM Quantum Canary Processors. The code base to perform all such simulations on the real hardware with and without error-mitigation schemes are implemented using Qiskit \cite{Qiskit}. Based on the analysis of the aforementioned results on quantum devices, it is evident that for good accuracy one needs to have an intuition of the starting step motivated by the physics of the problem. We use the ansatz as motivated in Section \ref{sec:ansatz_motivation} and warm-start the optimization by using initial parameters as discussed in the previous section. We fixed the choice of the optimizer as COBYLA since its performance is the best as discussed. Also, we performed the optimization using the two error mitigation schemes: Pauli-Twirling (T-Rex) and ZNE. The ground state energy obtained upon optimization using the above specifications is well within the desired accuracy of relative energy error 1$\%$ on the \textit{ibm$\_$hanoi} device. \textit{ibm$\_$hanoi} is a 27-qubit device with the median CNOT error being 7.504e-3, median readout error is 1.150e-2, median $T_1$ error of
$148.74 \:\mu s$, median $T_2$ error of
$116.04  \:\mu s$ as per the latest calibration statistics.

\begin{figure}[ht!]
  \centering
  \includegraphics[width=0.9\textwidth]{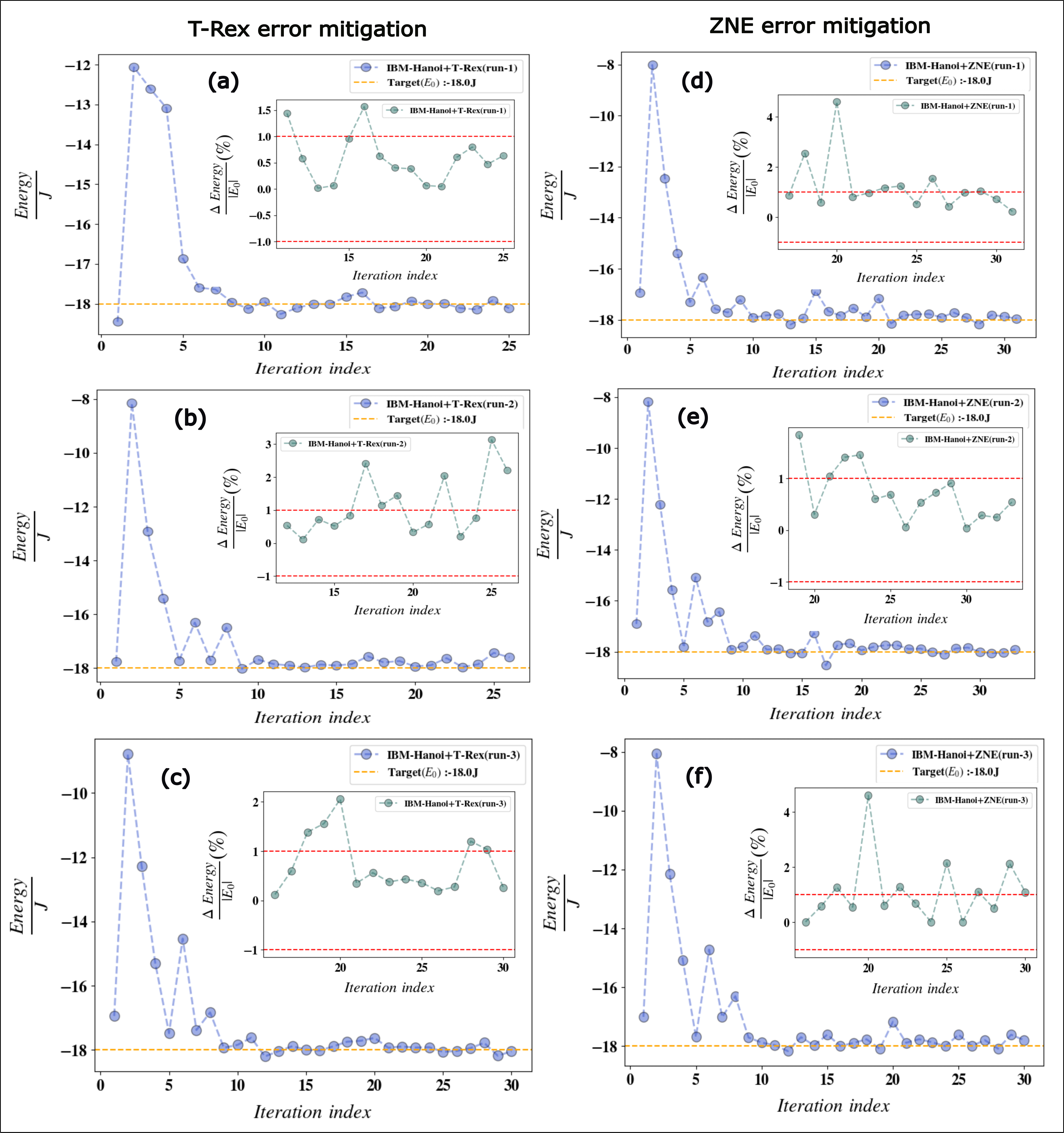}
  \caption{(a) Plots of energy obtained through optimization, as a function of the number of epochs for the full 12-qubit ansatz (See Fig.\ref{fig:ansatz_full}) using Pauli twirling (T-Rex) on \textit{ibm$\_$hanoi} quantum device (b) Same as in (a) with second instance of Pauli twirling (T-Rex) (c) Same as in (b) except with a third instance of Pauli twirling (T-Rex)(d) Plots of energy obtained through optimization, as a function of the number of epochs for the full 12-qubit ansatz (See Fig.\ref{fig:ansatz_full}) using zero-noise extrapolation(ZNE) on \textit{ibm$\_$hanoi}. (e) Same as in (d) with second instance of zero-noise extrapolation(ZNE) (f) Same as in (d) with third instance of zero-noise extrapolation(ZNE). Inset in each plot (a-f) shows the associated relative energy error $\%$ for each run over the last 15 iterations when a reasonable convergence is reached. The $\pm$ 1$\%$ relative energy error limit from the target energy $E_0= -18J$ is shown as a red-dashed window within each insets. For each instance we warm-start from an initial parameter set obtained by randomly sampling within $1\%$ error of the optimized parameter values $(\theta \sim \pi/2, \phi \sim \pi)$ procured from the ZNE-mitigated independent run of the two-qubit component as discussed in Fig.\ref{fig:2_qubit_all_pics}. COBYLA is used as the optimizer of choice for all cases.}
  \label{fig:Energy_trex_hanoi}
\end{figure}

We performed three runs using the T-Rex error mitigation scheme and three runs using the ZNE scheme (See Fig. \ref{fig:Energy_trex_hanoi}(a-f)). The results displayed in the figure shows the convergence of energy as a function of the epochs. As can be seen in the said figure, near the end of the optimization process the converged energy is very close to $-18J$ which is the desired value. Also, we show in each of the insets of Fig.\ref{fig:Energy_trex_hanoi}(a-f) the relative energy error computed using the
difference between the true energy $E_0=-18 J$ and the obtained ground state energy over the last few iterations, and the plots confirm that the accuracy of our optimization protocol is well within the relative error of 1$\%$ limit for most cases (the said limit is denoted using red dashed lines in insets of Fig.\ref{fig:Energy_trex_hanoi}(a-f)). We summarize the results in Table.\ref{tab:summary}
and also calculate the mean energy and the standard deviation using the last 15 iterations of the optimization process. The converged set of parameters ($\theta, \phi$) corresponds to the last iteration step, and the relative energy error $\%$ using the mean energy in the third column is also presented.
  
\begin{table}[h]
\centering
\Large
\resizebox{\textwidth}{!}{
\begin{tabular}{c|c|c|c|c|c}
\hline
\textbf{Device (Error mitigation scheme)} & \textbf{Job description} &\textbf{$E_{avg}$}/\textbf{J} & \textbf{$\sigma_{E_{avg}}$}/\textbf{J}  & \textbf{$\frac{|E_{avg}-E_0|}{|E_0|}\%$} & \textbf{Parameters($\bar{\theta} \pm \frac{\sigma_{\theta}}{2} , \bar{\phi} \pm \frac{\sigma_{\phi}}{2}$)}\\
\hline
IBM Hanoi (T-Rex) & run 1 (Fig.\ref{fig:Energy_trex_hanoi}(a)) & -18.0216  &  0.1307 &  0.12 & (1.607 $\pm 0.004$, 3.141 $\pm 0.003$)   \\
& run 2 (Fig.\ref{fig:Energy_trex_hanoi}(b)) & -17.7963 & 0.1591 &  1.13& (1.844 $\pm 0.004$, 3.083 $\pm 0.002$) \\
& run 3 (Fig.\ref{fig:Energy_trex_hanoi}(c)) & -17.9182  & 0.1427 & 0.45 & (1.509 $\pm 0.003$, 2.984 $\pm 0.004$)\\
\hline
IBM Hanoi (ZNE) & run 1 (Fig.\ref{fig:Energy_trex_hanoi}(d)) & -17.8233  &  0.2294 &  0.98 & (1.644 $\pm 0.004$, 3.092 $\pm 0.005$)  \\
& run 2 (Fig.\ref{fig:Energy_trex_hanoi}(e)) & -17.8996 & 0.1226 &  0.56 &(1.549 $\pm 0.002$, 2.974 $\pm 0.003$) \\
& run 3 (Fig.\ref{fig:Energy_trex_hanoi}(f)) & -17.8257  & 0.2267 & 0.97& (1.598 $\pm 0.003$, 3.236 $\pm 0.004$) \\
\hline 
\end{tabular}}
\caption{%
      Summary of results from actual hardware $ibm\_hanoi$. \textbf{$E_{avg}$}/\textbf{J} and \textbf{$\sigma_{E_{avg}}$}/\textbf{J} are the mean energy and the associated standard deviation respectively computed using the last 15 iterations of each run to account for fluctuations in self-convergence (See inset of Fig.\ref{fig:Energy_trex_hanoi}(a-f)). The relative energy error $\%$ in fifth column is computed using the $E_{avg}$ in the third column and the true energy value $E_0=-18 J$. The converged set of unique parameters ($\bar{\theta}, \bar{\phi}$) are also averaged over the same set of iterations and presented in radians along with their respective standard deviations ($\sigma_{\theta}, \sigma_{\phi}$) defining error ranges as $\pm \frac{\sigma_{\theta}}{2}$ and $\pm \frac{\sigma_{\phi}}{2}$. The exact values are ($\bar{\theta}=\frac{\pi}{2}, \bar{\phi}=\pi$) as discussed for noiseless simulations. }\label{tab:summary}
\end{table}

\section{Conclusion}
In this report we have shown the construction of a physically inspired ansatz for the quantum simulation of a resonance valence bond state (RVB) which is the ground state of a Kagome unit cell, commonly prescribed as a candidate for a gapped spin liquid. The ansatz discussed in this report heavily relies on the fact that the DMRG calculations on the said unit cell showed significant two-point correlations only for spins present on the sites in the outer triangles of a Kagome unit cell. Through suitable analysis, it was also shown that it is possible to construct a  ground state function which has a significant and negative correlation only for the adjacent sites on the outer triangles. Using the said information, we defined two auxillary hamiltonians with reduced measurements whose ground state is identical to the system of interest. Equipped with such an operator and an ansatz we performed a thorough analysis of the choice of optimizer required for a variational computation, the choice of initial state and the error-mitigation techniques to be employed. Lastly unlike in previous report\cite{PhysRevB.106.214429}, we validate all our inferences by showing result of computations within $1\%$ error in energy even on a real quantum device \textit{ibm$\_$hanoi}. 

\begin{figure}[htbp]
  \centering
  \includegraphics[width=0.5\textwidth]{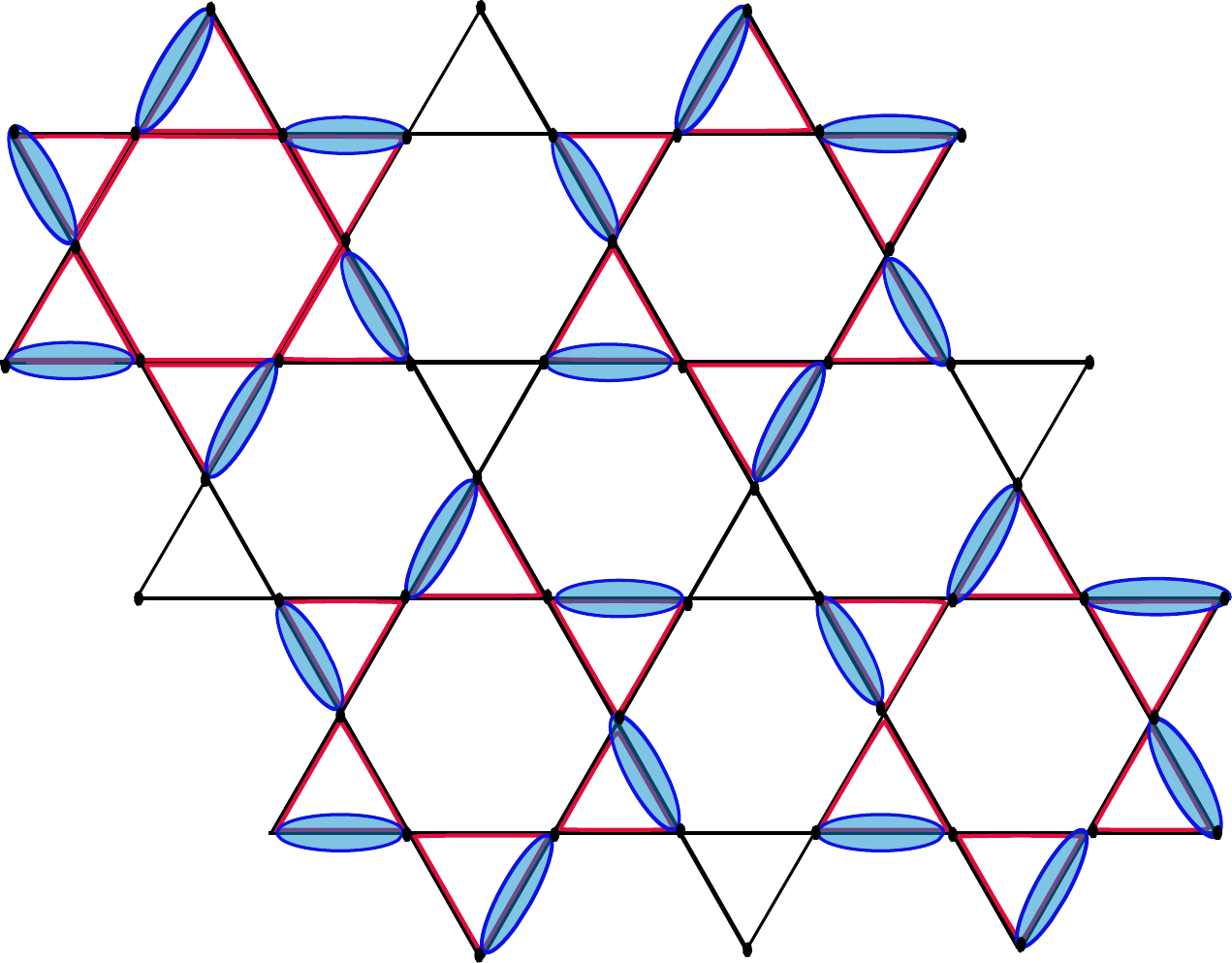}
  \caption{The figure shows the pinwheel structure of several Kagome unit cells. The ansatz that we picked satisfies the $C_6$ symmetry and also matches the spin correlations obtained for the ground state of this lattice\cite{kagomel} thereby making it extendable to handle larger Kagome lattices too}
  \label{fig: pin wheel}
\end{figure}

The ansatz we designed can be generalized to larger kagome lattices naturally. This could be attributed to the existence of long-range order in kagome Heisenberg anti-ferromagnets as discussed in \cite{PhysRevB.66.132408}. The authors of \cite{PhysRevB.66.132408} consider a kagome lattice as a set of stars with 12 spins arranged in a triangular lattice as in Fig.\ref{fig: pin wheel}. The authors show that the interaction between the stars leads to a band of singlet excitations and can be considered a perturbation in the low-energy sector. Extensive work exists in literature confirming the resonating valence Bond state nature of such ground states. \cite{Matan_2010, kagomel}. To quantify the scalability, consider a larger lattice with N qubits. Then, one can construct $N/12$ disjoint stars (Since each qubit is part of exactly one star.) So, our final ansatz would require $O(N/12)$ single and two-qubit gates at constant depth ($O(1)$). This is definitely advantageous compared to a generic ansatz (with larger depth) that doesn't exploit the hidden long-range order in the Kagome lattices or even the ansatz proposed in \cite{PhysRevB.106.214429} which uses a specialized $fSim$ gate implementable using a SWAP gate (requiring 3 CNOT gates on average for each bond) unlike us which requires just 1 CNOT gate per bond. There is also an advantage in terms of the 
unique parameter count (in our case it is just 2 for each unit cell) and also auxillary hamiltonians with reduced number of measurables (see Section \ref{sec:Aux_H}) and reduced number of shots thereby reducing statistical error. The simplification we used for the reduction in unique parameter count based on symmetry arguments can be adopted for other larger lattices with other symmetries depending on the topology of the lattice (For example, \cite{Yan_2011} looks at cylindrical and torus topology out of many).

Since our method entails preparing RVBs in real quantum devices with high fidelity, this holds significance across multiple areas of physics and chemistry 
even beyond the domain of quantum spin liquids. This is because the ground state preparation carried out in this paper would serve as a starting step for various studies one could conduct on systems that support RVBs as their ground states and be an impetus for areas in physical chemistry wherein they feature extensively. In physical chemistry RVBs arose fairly early due to Pauling in the theory of conjugated electrons in $\pi$-bonded organic frameworks. Thereafter it has been used to qualitatively understand bonding in graphene within lenghth scale of few atomic units \cite{PhysRevLett.107.086807} and also in explaining the electronic structure of recently synthesized  spiro-bis(1,9-disubstituted phenalenyl)boron based neutral radical conducting solids which despite having high density of states at the Fermi-level needs activated conductivity (with an activation energy of 0.054 eV) owing to the possible formation of a dimeric valence bonded ground state \cite{pal2005resonating, mandal2006resonating}. Charge density phases in the benzannulated variant due to interchain interaction thereby competing with the formation of the RVB state making the latter stable at certain temperatures have also been reported \cite{bag2010resonating}. It has also been used for explaining the breakdown of the free-electron Fermi liquid theory observed from Compton scattering experiments in lithium clusters from a quantum Monte Carlo based study which claims to show significant contribution to the cohesive energy of the cluster from electron pairing resulting in an RVB like state \cite{PhysRevB.79.035416}. Even in copper carbodiimide $(CuNCN)$, the anomalous temperature dependance of spin susceptibility has been attributed to formation of a possible RVB like state at low temperatures, a theory that enjoys semi-quantitative agreement with experiment \cite{tchougreeff2012structural}. In condensed matter physics, the earliest attention which RVBs enjoyed was in the celebrated work of Anderson and Sethna\cite{PhysRevB.35.8865, PhysRevB.37.627, anderson1987resonating, anderson1987resonating} to partially explain the phase diagram of traditional cuprates in high-Tc superconductivity. It was later extended by Kotliar\cite{PhysRevB.37.3664} to account for $s$-wave and $d$-wave superconducting parameters and thereafter by Baskaran to $MgB_2$ \cite{PhysRevB.65.212505}. Besides our circuit ansatz can also be used for the study of related valence bond solids with localized dimers like in the Mazumdar-Ghosh model \cite{majumdar1969next} with next to nearest neighbor interaction or under certain limiting conditions in higher-dimensional Shastry and Sutherland model\cite{shastry1981exact, miyahara1999exact}. Quite recently \cite{yang2021probing}, experimental creation of an RVB state in an artificial magnet has been reported with even single-site addressability in which $\rm{Ti}$ atom clusters are probed on a $\rm{MgO}$ surface using a scanning tunnelling microscopy showing evidence of dimerization. Also instead of dimerizing to form an overall singlet, dimerization into a triplet (called tRVB) have been proposed as the basic building block in ferromagnetic Strange metals like in Ref \cite{shen2020strange} and in certain transition metal dichalcogenides like 1-$TaS_2$ \cite{PhysRevB.105.075142}. In the light of these studies, we feel that probing the RVB structure on a quantum device should be of profound importance and a timely investigation that can benefit several disciplines. 



\section{Acknowledgements}
We express our sincere gratitude to Prof. Arnab Banerjee for engaging in thought-provoking discussions.
We acknowledge funding from DOE, Office of Science through the Quantum Science Center (QSC), a National Quantum Information Science Research Center and the
U.S. Department of Energy (DOE) (Office of Basic Energy Sciences) under award no. DE-SC0019215. We acknowledge the use of IBM Quantum services for this work. The views expressed are those of the authors, and do not reflect the official policy or position of IBM or the IBM Quantum team.
\providecommand{\latin}[1]{#1}
\makeatletter
\providecommand{\doi}
  {\begingroup\let\do\@makeother\dospecials
  \catcode`\{=1 \catcode`\}=2 \doi@aux}
\providecommand{\doi@aux}[1]{\endgroup\texttt{#1}}
\makeatother
\providecommand*\mcitethebibliography{\thebibliography}
\csname @ifundefined\endcsname{endmcitethebibliography}
  {\let\endmcitethebibliography\endthebibliography}{}

\end{document}